\documentclass{iaus}
\usepackage{graphicx}

\title[Migration of Comets to the Terrestrial Planets]
{Migration of Comets to the Terrestrial Planets}

\author[S.I. Ipatov and J.C. Mather]   
{Sergei I. Ipatov$^{1,2}$ and John C. Mather$^3$}%

\affiliation{$^1$ Department of Terrestrial Magnetism, Carnegie Institution of
Washington,
 5241 Broad Branch Road, N.W., Washington, D.C. 20015-1305, USA 
\break email: siipatov@hotmail.com 
\break http://www.dtm.ciw.edu/ipatov
\\[\affilskip]
$^2$ Space Research Institute, Moscow, Russia \\[\affilskip]
$^3$LASP, NASA/Goddard Space Flight Center, Greenbelt, USA
\break email: John.C.Mather@nasa.gov}

\pubyear{2006}
\volume{236}  
\pagerange{1--10}
\date{?? and in revised form ??}
\jname{Near Earth Objects, Our Celestial Neighbors: Opportunity and Risk}
\editors{A. Milani, G.B. Valsecchi \& D. Vokrouhlicky, eds.}
\begin{document}

\maketitle

\begin{abstract}
We studied the orbital evolution of Jupiter-family comets (JFCs),
Halley-type comets (HTCs), and long-period comets, and probabilities of their
collisions with planets. 
In our runs the probability of a collision of one object with the 
Earth could be greater than the sum of probabilities for thousands 
of other objects. Even without a contribution of such a few bodies, 
the  probability of a collision of a former JFC with the Earth 
was greater than 4$\cdot$10$^{-6}$. This probability is enough for 
delivery of all the water to Earth's oceans during formation of the 
giant planets. 
The ratios of probabilities of collisions of JFCs and HTCs with Venus and 
Mars to the mass of a planet usually were not smaller than that for Earth. 
Among 30,000 considered  objects with initial orbits close to those of JFCs,
a few objects got Earth-crossing orbits with semi-major axes 
$a$$<$2 AU and aphelion distances $Q$$<$4.2 AU, or even got 
inner-Earth ($Q$$<$0.983 AU), Aten, or typical asteroidal orbits, 
and moved in such orbits for more than 1 Myr (up to tens or even 
hundreds of Myrs). 
From a dynamical point of view, the fraction 
of extinct comets among near-Earth objects can exceed several tens of percent, 
but, probably, many extinct comets disintegrated into mini-comets 
and dust during a smaller part of their dynamical lifetimes if these 
lifetimes were large. 


\end{abstract}

\firstsection 
\section{Introduction}

      Farinella et al. (1993), Bottke et al. (2002), Binzel et al. (2002), 
and Weissman et al. (2002) believe that asteroids are the main source of 
near-Earth objects (NEOs). Wetherill (1988) supposed that half of NEOs are 
former short-period comets. Trans-Neptunian objects (TNOs) can migrate to 
the near-Earth space. Duncan et al. (1995) and Kuchner et al. (2002) 
investigated the migration of TNOs to Neptune's orbit, and Levison \& 
Duncan (1997) studied their migration from Neptune's orbit to Jupiter's orbit. 
Levison et al. (2006) studied formation of Encke-type objects. More references 
on papers devoted to the migration of bodies from different regions of the solar 
system to the near-Earth space can be found in our previous publications on 
this problem (Ipatov 1995, 1999, 2000, 2001, 2002; Ipatov \& Hahn 1999, 
Ipatov \& Mather 2003, 2004a-b, 2006a). As migration of TNOs to Jupiter's 
orbit was considered by several authors, Ipatov (2002), and Ipatov \& 
Mather (2003, 2004a-b, 2006a) paid particular attention to the orbital 
evolution of Jupiter-crossing objects (JCOs), considering 
a larger number of JCOs than before. 

In the present paper, we summarize our studies of migration of cometary objects into
NEO orbits, paying particular attention to the probabilities of collisions 
of cometary objects with the terrestrial planets. These studies are 
based on our previous runs and on some new runs. Earlier we did not consider
the evolution of orbits of Halley-type comets and long-period comets
and did not study the probabilities of collisions of different 
comets with the giant planets.
Though some used runs are the same as earlier, the discussion on 
migration of small bodies based on these runs is different.

\section{Initial data}

Ipatov \& Mather (2003, 2004a-b, 2006a) integrated the orbital 
evolution of about 30,000 objects with initial orbits close to that
of Jupiter-family comets (JFCs). We considered  
the gravitational influence of planets, but omitted the influence 
of Mercury (except for Comet 2P) and Pluto. In about a half of 
runs we used the method by Bulirsh-Stoer (1966) (BULSTO code), 
and in other runs we used a symplectic method (RMVS3 code). The 
integration package of Levison \& Duncan (1994) was used.  

      In the first series of runs (denoted as $n1$) we calculated 
the orbital evolution of 3100 JCOs moving in initial 
orbits close to those of 20 real comets (with numbers 7, 9, 10, 
11, 14, 16, 17, 19, 22, 26, 30, 44, 47, 
51, 57, 61, 65, 71, 73, and 75) with period 5$<$$P_a$$<$9 yr, and in 
the second series of runs (denoted as $n2$) 
we considered 13,500 JCOs moving in initial orbits close to 
those of 10 real comets (with numbers 77, 81, 
82, 88, 90, 94, 96, 97, 110, and 113) with period 5$<$$P_a$$<$15 yr. 
In other series of runs, initial orbits 
were close to those of a single comet 
(2P/Encke, 9P/Tempel 1, 10P/Tempel 2, 22P/Kopff, 
28P/Neujmin 1, 39P/Oterma, or 44P/Reinmuth 2). 
In order to compare the orbital evolution of comets and asteroids, 
we also studied the orbital evolution of 1300 
asteroids initially moving in the 3:1 and 5:2 resonances with Jupiter. 

In our recent runs we also considered objects 
started from orbits of test long-period comets ($e_o$=0.995, 
$q_o$=$a_o$$\cdot$(1-$e_o$)=0.9 AU or 
$q_o$=0.1 AU, initial inclination $i_o$ varied from 0 to 180$^\circ$ 
in each run, bodies started at perihelion; 
these runs are denoted as $lpc$ runs) and test Halley-type comets ($a_o$=20 AU,
$i_o$ varied from 0 to 180$^\circ$ in each run, 
bodies started at perihelion; in some runs $e_o$=0.975 and 
$q_o$=0.5 AU, in other runs 
$e_o$=0.9 and $q_o$=2 AU; these runs are denoted as $htc$ runs). 

Usually we investigated the orbital evolution during the dynamical 
lifetimes of objects (at least until all the objects reached perihelion 
distance $q$$>$6 AU or collided with the Sun). Ipatov et al. (2004) 
and Ipatov \& Mather (2006a) studied migration of dust particles 
started from several comets, asteroids, and trans-Neptunian objects.

In our runs, planets were considered as material points, so literal 
collisions did not occur. However, using the algorithm suggested by 
Ipatov (2000) with the correction that takes into account a different 
velocity at different parts of the orbit (Ipatov \& Mather 2003), 
and based on all orbital elements sampled with a 500 yr step, we 
calculated the mean probability $P$ of collisions of
migrating objects with a planet. We define $P$ 
as $P_{\Sigma}/N$, where $P_{\Sigma}$ is the probability for 
all $N$ objects of a collision of an object with a planet 
during an object's dynamical lifetime.
Note that our algorithm differed from the \"Opik's scheme, and
included calculations of a synodic period and the region
where the distance between the `first' orbit and the projection
of the `second' orbit onto the plane of the `first' orbit
is less than the sphere of action (i.e., the Tisserand sphere).

For BULSTO runs, the integration step error was less than $\epsilon$, 
where $\epsilon$ varied between $10^{-13}$ and $10^{-8}$ (most of 
the runs were made for $\epsilon$ equal to $10^{-8}$ and $10^{-9}$), 
and for a RMVS3 runs an integration step $d_s$ varied from 0.1 to 
30 days (most runs were made for $d_s$=10 days). In a single run
with $N$ (usually $N$=250) objects, 
$\epsilon$ or $d_s$ was constant.  
Results obtained with the use of different methods of integration 
and different integration step were similar, exclusive for probabilities 
of collisions with the Sun in such runs when this probability was large 
(for Comet 2P, Comet 96P from $n2$ series, and the 3:1 resonance with 
Jupiter). Probabilities of collisions of bodies with planets 
were close for different integrators
even in the latter case because soon after close encounters 
with the Sun, bodies were ejected into hyperbolic orbits or 
moved in highly inclined orbits.

      H. Levison showed that it is difficult
to detect solar collisions in any numerical integrator, so he
removed objects with perihelion distance $q$$<$$q_{\min}$.
Our runs were made for direct 
modeling of collisions with the Sun, but we studied what happens if we 
consider $q_{\min}$ equal to $k_S$ radii $r_S$ of the Sun.  
We obtained that the mean probabilities of 
collisions of bodies with planets, lifetimes of objects that spent millions 
of years in Earth-crossing orbits, and other obtained results 
were practically the same if we consider that objects disappear 
when  $q$ becomes less than the radius $r_S$ of the Sun 
or even several such radii (i.e., we checked $q$$<$$k_S$$\cdot$$r_S$, 
where $k_S$ equals 1, 2, or another value). The only noticeable difference 
was for Comet 96P from $n2$ series and a smaller one was for Comet 2P,
but results of such runs were not included in our statistics. 
Eccentricity and inclination of Comet 96P/Machholz 1 are large, so usually even 
for these $n2$ runs the collision probabilities of objects with the terrestrial 
planets were not differed by more than 15\% at $k_S$=0 and  $k_S$=1. 
Among more than a hundred considered runs, 
there were three runs, for each of which at $k_S$=0 a body in an orbit close 
to that of Comet 96P was responsible for 70-75\% of collision probabilities  
with the Earth, and for $k_S$=1 a lifetime of such body was much less than 
for $k_S$=0. Nevertheless, for all ($\sim$$10^4$) objects from $n2$ series, 
at $k_S$=0 the probabilities of collisions with the terrestrial planets were 
close to those at $k_S$=1, even if we consider the above runs. The difference 
for times spent in Earth-crossing orbits is 
greater than that for the probabilities and is about 20\%. For all runs at 2P series, 
the difference in time spent in orbits with  $Q$$<$4.7 AU for $k_S$=0 and for $k_S$=1 
was less than 4\%. In 2P series of runs (and also for the 3:1 resonance with Jupiter), at 
$k_S$=0 we sometimes got orbits with $i$$>$$90^\circ$, but practically 
there were no such orbits at $k_S$$\ge$1 (Ipatov \& Mather 2004a-b). 
Among objects with initial orbits close to that of Comet 96P,
 we found one object which also got $i$$>$$90^\circ$ for 3 Myr. 
Inclinations of other such objects 
did not exceed 90$^\circ$.

         In most `cometary' runs (exclusive for 2P runs),
the fraction $P_{Sun}$ of comets collided with the Sun was less than 0.02; 
$P_{Sun}$ exceeded 5\% for some $htc$ runs, and most of comets in 2P runs
collided with the Sun.

\section{Computer simulations of migration of comets into near-Earth object orbits}

Some migrating JCOs got Earth-crossing orbits. Usually they spent 
in such orbits only a few thousands of years, but a few considered objects 
moved in Earth-crossing orbits with aphelion distances $Q$$<$4.2 AU for 
millions of years. The total times which 30,000 considered objects started
from JFC orbits  spent  in Earth-crossing 
orbits with $a$$<$2 AU were due to a few tens of objects, but
mainly only to a few of them. With BULSTO 
at $10^{-9}$$\le$$\epsilon$$\le$$10^{-8}$, six and nine objects, 
respectively from 10P and 2P series, moved into Apollo orbits with 
$a$$<$2 AU ($Al2$ orbits) for at least 0.5 Myr each, and five of them 
remained in such orbits for more than 5 Myr each. Among the JFCs 
considered with BULSTO, only one and two JFCs reached inner-Earth orbits 
(IEO, $Q$$<$0.983 AU) and Aten orbits, respectively. Only two objects in 
series $n2$ got $Al2$ orbits during more than 1 Myr.
For the $n1$ series 
of runs, while moving in JCO orbits, objects had orbital periods 
$P_a$$<$20 yr (Jupiter-family comets) and 20$<$$P_a$$<$200 yr 
(Halley-type comets) for 32\% and 38\% of the mean value $T_J$ ($T_J$=0.12 Myr)
of the total time spent by one object in 
Jupiter-crossing orbits, respectively.

Four considered former JFCs even got IEO 
or Aten orbits for Myrs. Note that Ipatov (1995) 
obtained migration of JCOs into IEO orbits using the method of spheres to 
consider the gravitational influence of planets.
In our BULSTO runs, one former JCO, which had an initial orbit close 
to that of 10P, moved in Aten orbits for 3.45 Myr, and the probability 
of its collision with the Earth from such orbits was 0.344. It also 
moved for about 10 Myr in IEO orbits before its collision with 
Venus, and during this time the probability of its collision with Venus 
was $P_V$=0.655. The above probabilities are greater than the total 
probabilities for $10^4$ other JCOs. Another object (from 2P BULSTO run) 
moved in highly eccentric Aten orbits for 83 Myr, and its lifetime before 
collision with the Sun was 352 Myr. Its probability of collisions with Earth, 
Venus, and Mars during its lifetime was 0.172, 0.224, and 0.065, respectively. 
With RMVS3 at $d_s$$\le$10 days for 2P run, the probability of 
collisions with Earth for one 
object was greater by a factor of 30 than for 250 other objects. 
For series $n1$ with RMVS3, the probability of a collision with the Earth for 
one object with an
initial orbit close to that of Comet 44P was 88\% of the total probability 
for 1200 objects from this series, and the total probability for 1198 
objects was only 4\%. For series 44P with $N$=1500 there were no objects with 
$a$$<$2 AU and $q$$<$1 AU, though the 44P object in $n1$ run spent 11.7 Myr 
in such orbits. For series $n2$ with RMVS3, we obtained one object with an initial 
orbit close to that of Comet 113P/Spitaler with relatively large values of probabilities 
of collisions with Earth and Venus. This object is responsible for 10\% of the 
total collision probability with Earth for all $n2$ objects, but 
most of the time spent by all these objects in orbits with $a$$<$2 AU and $q$$<$1 AU
are due to this object. Though about 
a half of 30,000 considered objects belong to series $n2$, most of objects that spent a 
long time in Earth-crossing orbits with $Q$$<$4.2 AU belong to other series of runs.

After 40 Myr one considered object with an initial orbit close to that of Comet 88P/Howell 
(from $n2$ RMVS3 runs) got  $Q$$<$3.5 AU, and it moved in orbits with  
$a$=2.60-2.61 AU,  1.7$<$$q$$<$2.2 AU, 3.1$<$$Q$$<$3.5 AU, eccentricity $e$=0.2-0.3, 
and inclination $i$=5-10$^\circ$ for 650 Myr. Another object (with an initial orbit 
close to that of Comet 94P/Russel 4) moved in orbits with $a$=1.95-2.1 AU, $q$$>$1.4 AU, 
$Q$$<$2.6 AU, $e$=0.2-0.3, and $i$=9-33$^\circ$ for 8 Myr (and it had $Q$$<$3 AU 
for 100 Myr). So JFCs can very rarely get typical asteroid orbits and move in them for 
Myrs. In our opinion, it can be possible that Comet 133P (Elst-Pizarro) moving in 
a typical asteroidal orbit (Hsieh \& Jewitt 2006)
was earlier a JFC and it circulated its orbit also 
due to non-gravitational forces. 
JFCs got typical asteroidal orbits less often than NEO orbits. 

Levison et al. (2006) argued that our obtained orbits with $a$$\approx$1 AU 
were due to the fact that collisions of objects with terrestrial planets were not 
taken into account in our runs and such orbits were caused by too close
encounters of objects with planets which really result in collisions. 
Based on the orbital elements obtained 
in our runs, we can conclude that probabilities of collisions of migrating 
bodies with planets before bodies got orbits with $a$$<$2 AU were very small 
and the reason of the transformations of orbits was not caused by such close encounters 
of objects with the terrestrial planets 
that really resulted in collisions with the planets. 
Some real probabilities of collisions of bodies 
moving in orbits with $a$$<$2 AU with planets 
were only after bodies had already got such 
orbits and moved in them for tens or hundreds of Myr.
Other scientists did not obtain the migration of JCOs into 
orbits with $a$$\approx$1 AU because they considered
other initial data.
In series $n2$ with 13,500 objects, we also did not obtain orbits
with $a$$\approx$1 AU and obtained only two orbits with $a$$<$2 AU
(the latter orbits were also obtained by Levison et al. 2006).
For other series of runs, we paid more attention to
those initial data for which migrating objects could spend a long time
inside Jupiter's orbit. 

\section{Cometary objects in NEO orbits}
 
Based on the results of migration of JFCs with initial 
orbits close to the orbit of Comet P/1996 R2 obtained by Ipatov \& Hahn (1999) 
(for these runs  with about a hundred objects, there were no objects which  
spent a long time in Earth-crossing orbits), Ipatov (1999, 2001) found that 
10-20\% or more of all 1-km Earth-crossers could have come from the 
Edgeworth-Kuiper belt into Jupiter-crossing orbits. 
Using our results of the orbital evolution of 30,000 JCOs and the results of 
migration of TNOs obtained by Duncan et al. (1995) and 
considering the total of 5$\cdot$$10^9$ 1-km TNOs with  30$<$$a$$<$50 AU, 
Ipatov \& Mather (2003, 2004a-b) estimated the number of 1-km former 
TNOs in NEO orbits. 
Results of their runs testify in favor of at least one of these conclusions: 
1) the portion of 1-km former TNOs among NEOs can exceed several tens of percents, 
2) the number of TNOs migrating inside the solar system could be smaller by a 
factor of several than it was earlier considered, 3) most of 1-km former TNOs 
that had got NEO orbits disintegrated into mini-comets and dust during a smaller 
part of their dynamical lifetimes if these lifetimes are not small. 
All these three factors could take place.
We consider that the role of disintegration may be more valuable and most 
of former comets that could move inside Jupiter's orbit for millions 
of years really were disintegrated.
As the number of TNOs, 
their rate of migration inside the solar system, 
and lifetimes of former comets before their disruption are not well known, 
the estimates of the fraction of former TNOs among NEOs are very approximate.

Disintegrated comets could produce a lot of mini-comets and dust.
Therefore there could be a lot of cometary dust 
among zodiacal 
particles, some of them were produced by high eccentricity 
comets (such as Comet 2P/Encke). The same conclusion about cometary dust was made by 
Ipatov et al. (2006a-b) based on analysis of spectra of the zodiacal light.
Dynamical lifetimes of dust particles are typically smaller than those
for bodies and they are smaller for smaller particles.
So old extinct comets are not surrounded by dust.
Frank et al. (1986a-b) concluded that there is a large influx of
small comets into the Earth's upper atmosphere.

It is known (Merline et al. 2002, Noll 2006, Pravec et al. 2006)
that about 15\% of NEOs and 2-3\% of main-belt asteroids are binaries.
We can suppose that the fraction of NEO binaries is greater for those NEOs 
which are extinct comets than for asteroidal NEOs. Comets more often
split into smaller parts than asteroids, and probably there are 
former comets even among binary main-belt asterods.
Besides, if initial (before collisional destruction) small bodies were formed by  
compression of dust condensations, then the fraction of binary objects
is greater for greater distances of the place of origin of bodies
from the Sun (Ipatov 2004).  

      Comets are estimated to be active for $T_{act}$$\sim$10$^3$-10$^4$ yr. 
$T_{act}$ is smaller for closer encounters with the Sun (Weissman et al., 2002), 
so for Comet 2P it is smaller than for other JFCs. 
If considered as material points, some former comets can move for tens or 
even hundreds of Myr in NEO orbits, so the number of extinct comets can 
exceed the number of active comets by several orders of magnitude. The mean 
time spent by Encke-type objects in Earth-crossing orbits was $\ge$0.4 Myr. 
This time corresponds to $\ge$40-400 extinct comets of this type if we
consider that Encke-type active comet is not  an exceptional 
event in the history of the solar system. Note that 
the diameter of Comet 2P is about 5-10 km, so the number of 1-km Earth-crossing 
extinct Encke-type comets can be greater by a factor of 25-100 than 
the above estimate for Encke-size comets and can exceed 1000 for such estimates.
The rate of a cometary object 
decoupling from the Jupiter vicinity and transferring to a NEO-like orbit 
can be increased by a factor of several due to nongravitational effects 
(Harris \& Bailey 1998, Asher et al. 2001, Fernandez \& Gallardo 2002). 
The role of the Yarkovsky and YORP effects on dynamics of asteroids
was summarized by Bottke et al. (2006).


Dynamical models of the NEO population considered by Bottke et al. (2002)
allowed 6 \% of dead comets.
From measured albedos, Fernandez et al. (2001) concluded that the fraction 
of extinct comets among NEOs and unusual asteroids is significant (9\%). 
Rickman et al. (2001) and Jewitt \& Fernandez (2001) considered that dark 
spectral classes that might include the ex-comets are severely 
under-represented and comets played an important and perhaps even 
dominant role among all km-size Earth impactors.
Binzel \& Lupishko (2006) studied albedos and spectra of NEOs
and concluded that 15$\pm$5 \% of the entire NEO population
may be comprised by extinct or dormant comets.
Harris \& Bailey (1998) concluded that the number of cometary
asteroids becomes comparable to the
number of bodies injected from the main asteroid belt
if one considers non-gravitational effects.
Typical comets have larger rotation periods than typical NEOs 
(Binzel et al. 1992, Lupishko \& Lupishko 2001), but, while losing 
considerable portions of their masses, extinct comets can decrease 
these periods.
            
     Our runs showed that if one observes former comets in NEO orbits, then 
most of them could have already moved in such orbits for millions 
(or at least hundreds of thousands) of years. 
Some former comets that have moved in typical NEO orbits for millions 
of years, and might have had multiple close 
encounters with the Sun, could have lost their mantles, which caused their 
low albedo, and so change their albedo (for most observed NEOs, the albedo 
is greater than that for comets (Fernandez et al. 2001)) and would look like 
typical asteroids.
For better estimates of the portion of extinct comets among NEOs we will need 
orbit integrations for many more TNOs and JCOs, and wider analysis of 
observations and craters. 
 

\section{Probabilities of collisions of comets with planets}
The probability of a collision of one celestial body with a planet can be greater 
than the total probability for thousands of objects with almost the same initial
orbit. A few JCOs (mentioned in Section 3)
with the highest probabilities with planets were not 
included in the statistics presented below. For series $n1$, the 
probability $P_E$ of a collision of an object with the Earth 
(during a dynamical lifetime of the object) was about 4.5$\cdot$$10^{-6}$ 
and 4.8$\cdot$$10^{-6}$ for BULSTO and RMVS3 runs, respectively (but for RMVS3 it is by an 
order of magnitude greater if we consider one more object with the highest probability). 
For series $n2$, the mean value of $P_E$ was $\sim$(10-15)$\cdot$$10^{-6}$ 
for BULSTO and RMVS3 runs.

Probabilities of collisions of JFCs with planets were different for different comets. 
The probability of a collision of Comet 10P with the Earth 
was 36$\cdot$$10^{-6}$ and 22$\cdot$$10^{-6}$ 
for BULSTO and RMVS3 runs, respectively ($P_E$=140$\cdot$$10^{-6}$ 
if we include objects with high collision probabilities).
For 2P runs, $P_E$ was relatively large: $\approx$(1-5)$\cdot$$10^{-4}$. 
For most other considered JFCs, $10^{-6}$$\le$$P_E$$\le$$10^{-5}$.
 For Comets 22P and 39P, $P_E$$\approx$(1-2)$\cdot 10^{-6}$, and for Comets 
9P, 28P and 44P, $P_E$$\approx$(2-5)$\cdot 10^{-6}$. 
The Bulirsh-Stoer method of integration and a symplectic method gave similar results. 
    Values of $P_E$ were about (0.5-2)$\cdot$$10^{-6}$ for $htc$ runs, 
with greater values for smaller $q_o$.
For $lpc$ runs, $P_E$=0.6$\cdot$$10^{-6}$ at $q_o$=0.9 AU and 
$P_E$=0.25$\cdot$$10^{-6}$ at $q_o$=0.1 AU.
Dynamical lifetimes of some objects in $htc$ and $lpc$ runs
exceeded several Myr.
Note that we considered collision probabilities for
objects starting from different types of orbits, but
a type of orbit (e.g. JFCs, HTCs, and LPCs) can change
during the orbital evolution of objects.

    The fraction of asteroids migrated from the 3:1 resonance with Jupiter 
that collided with the Earth was greater by a factor of several than that 
for the 5:2 resonance ($P_E$$\sim$$10^{-3}$ and $P_E$$\sim$(1-3)$\cdot$$10^{-4}$, 
respectively). The probabilities of collisions with the Earth for resonant 
asteroids (per one object) were about two orders of magnitude greater than 
those for typical JFCs. This difference in the probabilities is greater for the asteroids 
and TNOs, as not all TNOs that had leaved the trans-Neptunian belt reached Jupiter's orbit. 
The present mass of the Edgeworth-Kuiper belt is considered to be about 
two orders of magnitude greater than that of the main asteroid belt.
      For dust particles started from comets and asteroids, $P_E$  was maximum for 
diameters $d$$\sim$100 $\mu$m (Ipatov et al. 2004, Ipatov \& Mather 2006a-b). 
These maximum values of $P_E$   were usually 
(exclusive for 2P runs) greater at least by an order of magnitude than the values 
for parent comets.

The probabilities $P_V$ of collisions of JFCs and HTCs with Venus usually did not 
differ by more than a factor of 2 from those with Earth. For 2P runs, they 
were greater than those with Earth, but in most of other runs they were smaller. 
The probabilities $P_M$ of collisions of JFCs and HTCs with Mars usually were smaller by 
a factor of 3-6  (10 for 2P runs) than those with Earth, i.e., 
Mars accreted more cometary bodies 
than Earth per unit of a mass of a planet.
For $lpc$ runs, the values of $P_E$ and $P_V$
can differ by a factor of 3, and $P_E/P_M$$\sim$7-10.

For most our runs, the probability $P_J$ of a collision of a JFC 
with Jupiter (during a dynamical lifetime of the comet) was 
$\sim$0.01. Usually it was less than 0.03, though
it can be up to 0.06 in a single run. 
The mean time $T_J$ spent by
bodies in Jupiter-crossing orbits was 0.12 Myr for $n1$ runs.
Therefore a collisional lifetime of a hypothetical object
in Jupiter-crossing orbit can be estimated as 10 Myr
for $n1$ and $n2$ runs, and it was much greater for comets in highly
eccentric orbits. As considered bodies never spent such long times in Jupiter-crossing 
orbit, it may be more correct to note that the collision frequency
of objects starting from JFC orbits and moving in Jupiter-crossing orbits 
is about 10$^{-7}$ yr$^{-1}$.
Though $T_J$ can be a little greater for 2P runs
than for $n1$ and $n2$ runs, and it can exceed 1 Myr for $htc$ runs, 
$P_J$ was only about 5$\cdot$$10^{-4}$ for some 2P and $htc$ runs. 
In other 2P runs, $P_J$ can be  greater or smaller by a factor of 20
than the above value. 
For $lpc$ runs, $P_J$ was smaller by an order of magnitude than that for
$htc$ runs though $T_J$ did not differ much.
Probabilities $P_S$ of collisions of bodies from $n1$ and $n2$ runs
with Saturn typically were smaller by an order of magnutude than those with Jupiter, and
collision probabilities with Uranus and Neptune typically were smaller by 
three orders of magnitude than those with Jupiter.
The ratio of probabilities of collisions of bodies
with different giant planets, for a pair of planets can vary
by more than an order of magnitude from run to run.

As only a small fraction of comets collided with planets during dynamical lifetimes
of comets, the orbital evolution of comets for the considered model 
of material points was close to that for the model when comets collided with a planet 
are removed from integrations.

\section{Delivery of water and volitiles to planets}

Using $P_E$=$4\cdot10^{-6}$ (this value is smaller than the mean value of $P_E$
obtained in our runs for JFCs) and assuming that 
the total mass of planetesimals that ever crossed Jupiter's orbit 
is  about 100$m_{\oplus}$ (Ipatov 1987, 1993), 
where $m_{\oplus}$ is the mass of the Earth,  we obtain that the total mass of water 
delivered from the feeding zone of the giant planets to the Earth could be about 
the mass of water in Earth's oceans. 
We considered that the fraction $k_w$ of water in planetesimals equaled 0.5.
For present comets $k_w$$<$0.5 (Jewitt 2004), but it is considered that
$k_w$ could  exceed 0.5 for planetesimals.
The fraction of the mass of the planet 
delivered by JFCs and HTCs can be greater for Mars and Venus than that 
for the Earth. This larger mass fraction would result in relatively 
large ancient oceans on Mars and Venus.
The conclusion that planetesimals from the zone of the giant planets
could  deliver all the water to the terrestrial oceans was also made by Ipatov (2001)
and Marov \& Ipatov (2001) on the basis of runs by Ipatov \& Hahn (1999).
      
      The above estimate of water delivery by cometary bodies
to the Earth is greater than those by 
Morbidelli et al. (2000) and Levison et al. (2001), but is in accordance with the results by 
Chyba (1989) and Rickman et al. (2001). The larger value of $P_E$ we have 
calculated compared to those argued by Morbidelli et al. (2000) 
($P_E$$\sim$(1-3)$\cdot$$10^{-6}$) 
and Levison et al. (2001) ($P_E$=4$\cdot$$10^{-7}$) is caused by the fact that 
in our runs we considered different initial orbits and a larger number of JCOs.
Levison et al. (2001) 
did not take into account the influence of the terrestrial planets, so probably 
that is why his values of $P_E$ are even smaller than those by Morbidelli et al. (2000). 
The latter authors used results of integrations of objects initially located
beyond Jupiter's orbit. For 39P runs ($a_o$=7.25 AU and $e_o$=0.25), 
we obtained $P_E$ equal to 1.2$\cdot$$10^{-6}$ and 2.5$\cdot$$10^{-6}$ for 
BULSTO and RMVS3 runs, respectively. These values are in accordance with
the values of $P_E$ obtained by Morbidelli et al.
Morbidelli et al. (2000) considered reasonable that about 50-100$m_{\oplus}$ 
of planetesimals primordially existed in the Jupiter-Saturn region and about 
20-30$m_{\oplus}$ of planetesimals in the Uranus-Neptune region. We think 
that they considerably underestimated the mass of 
planetesimals in the Uranus-Neptune region.

Lunine (2004, 2006) concluded that possible sources of water for Earth 
are diverse, and include Mars-sized hydrated bodies in the asteroid belt, 
smaller ``asteroidal'' bodies, water adsorbed into dry silicate grains 
in the nebula, and comets.
Lunine et al. (2003) considered  most of the Earth's water as a product
of collisions between the growing Earth and planet-sized ``embryos''
from the asteroid belt.
Drake \& Campins (2006) noted that the key argument against an asteroidal
source of Earth's water is that the O's isotopic composition of Earth's primitive 
upper mantle matches that of anhydrous ordinary chondrites, not
hydrous carbonaceous chondrites.
Kuchner et al. (2004) investigated the possibility that the Earth's ocean water
originated as ice grains formed in a cold nebula,
delivered to the Earth by drag forces from co-orbital nebular gas.
Dust particles could also deliver water to the Earth from the feeding
zone of the giant planets. Ipatov \& Mather (2006a,b) obtained
that the probability of collisions of 10-100 $\mu$m particles
originated beyond Jupiter's orbit is about (1-3)$\cdot$$10^{-4}$.
Therefore the water in the terrestrial oceans (2$\cdot$$10^{-4}$$m_\oplus$)
can be delivered by particles (for the model without sublimation) which had
contained $\sim$$m_\oplus$ of water when they had been located beyond Jupiter.
So dust particles could also play some role in the delivery of water
to the terrestrial planets during planet formation.

      There is the deuterium/hydrogen paradox of Earth's oceans 
(D/H ratio is different for oceans and 
comets), but Pavlov et al. (1999) suggested that solar wind-implanted hydrogen 
on interplanetary dust particles provided the necessary 
low-D/H component of Earth's water inventory, 
and Delsemme (1999) considered that most of the seawater 
was brought by the comets that originated 
in Jupiter's zone, where steam from the inner solar system condensed 
onto icy interstellar grains before they accreted into larger bodies.
It is likely (Drake \& Campins 2006) that the D/H and Ar/O
ratios measured in cometary comas and tails are not truly
representative of cometary interiors.

Small bodies which collided with planets could deliver volatiles 
and organic/prebiotic compounds needed for life origin.
     Marov \& Ipatov (2005) concluded that
dust particles could be most efficient in the delivery of organic
or even biogenic matter to the Earth, because they experience substantially
weaker heating when passing through the atmosphere (an excess heat
is radiated effectively due to high total surface-to-mass
ratio for dust particles). They assumed that life forms drastically
different from the terrestrial analogs
are unlikely to be found elsewhere in the Solar System
(if any), e.g., either extinct or extant life on Mars.

\section{Conclusions}

      Some Jupiter-family comets can reach typical NEO orbits and 
remain there for millions of years. 
From the dynamical point of 
view (if comets didn't disintegrate) there could be 
(not 'must be') many (up to tens of percent) extinct comets among the NEOs,
but, probably, many extinct comets disintegrated into mini-comets 
and dust during a smaller part of their dynamical lifetimes if these 
lifetimes were large. Disintegration of comets 
can provide a considerable fraction of cometary dust among 
the zodiacal dust particles.
The probability of a collision of one object 
moving for a long time in Earth-crossing orbits, with the 
Earth could be greater than the sum of probabilities for thousands 
of other objects, even having similar initial orbits. 
Even without a contribution of such a few bodies, 
the probability of a collision of a former JFC
(during its dynamical lifetime) with the Earth 
was greater than 4$\cdot$10$^{-6}$. This probability is enough for 
delivery of all the water to Earth's oceans during formation of the 
giant planets. 
The ratios of probabilities of collisions of JFCs and HTCs with Venus and 
Mars to the mass of a planet usually were not smaller than that for Earth. 



\begin{thebibliography}{}

\bibitem[Asher et al. (2002)]{Asher2002} {Asher, D.J., Bailey, M.E., \& Steel, D.I.}  2001, 
in: M. Ya. Marov \& H. Rickman (eds.),
      \textit{Collisional Processes in the Solar System}, ASSL, vol. 261, p. 121
      
\bibitem[]{}{Binzel, R.P., Xu, S., Bus, S.J., \& Bowell, E.} 1992, 
      \textit{Science} 257, 779 
      
 \bibitem[]{}      
Binzel, R.P., Lupishko, D.F., Di Martino, M., et al. 2002, 
in: W.F. Bottke Jr., A. Cellino, P. Paolicchi, \& R.P. Binzel (eds.),
\textit{Asteroids III}, Univ. of Arizona: Tucson, p. 255 

 \bibitem[]{}      
Binzel, R.P. \& Lupishko, D.F. 2006, 
in: D. Lazzaro, S. Ferraz-Mello, \& J.A. Fernandez (eds.),
\textit{Asteroids, Comets, and Meteors}, IAU Symposium 229, 
Cambridge University Press: Cambridge, p. 207 

\bibitem[Bottke et al. (2002)]{Bottke2002} Bottke, W.F., Morbidelli, A., Jedicke, R.,  
Petit, J.M., Levison, H.F., Michel, P., \& Metcalfe, T.S. 2002, \textit{Icarus} 156, 399

\bibitem[]{} Bottke, W.F., Vokrouhlicky, D., Rubincam, D.P.,
\& Nesvorny, D. 2006, \textit{Annu. Rev. Earth Planet. Sci.} 34, 157

 \bibitem[]{}  Bulirsh, R. \& Stoer, J. 1966, 
\textit{Numer. Math.} 8, 1

 \bibitem[]{} Chyba, C.F. 1989, 
\textit{Nature} 343, 129 
      
 \bibitem[]{} Delsemme, A.H. 1999, 
\textit{Planetary \& Space Science} 47, 125

 \bibitem[]{}      
Drake, M. \& Campins, H., 2006, 
in: D. Lazzaro, S. Ferraz-Mello, \& J.A. Fernandez (ed.),
\textit{Asteroids, Comets, and Meteors}, IAU Symposium 229, 
Cambridge University Press: Cambridge, p. 381 

\bibitem[Duncan et al. (1995)]{Duncan1995} {Duncan, M.J.,  Levison, H.F., 
\&  Budd, S.M.} 1995, \textit{Astron. J.} 110, 3073

\bibitem[Farinella et al. (1993)]{Farinella1993} {Farinella, P., Gonczi, R., Froeschle, Ch., \&
Froeschle, C.} 1993, \textit{Icarus} 101, 174

 \bibitem[]{} Fernandez, J.A. \& Gallardo, T. 2002, 
\textit{Icarus} 159, 358
      
  \bibitem[]{}  Fernandez, Y.R., Jewitt, D.C., \& Sheppard, S.S. 2001, 
\textit{ApJ} (Letters) 553, L197

  \bibitem[]{}  Frank, L.A., Sigwarth, J.B., \& Graven, J.D.  1986a,
\textit{Geophys. Res. Lett.} 13, 303

  \bibitem[]{}  Frank, L.A., Sigwarth, J.B., \& Graven, J.D.  1986b,
\textit{Geophys. Res. Lett.} 13, 307

  \bibitem[]{}  Harris, N.W. \& Bailey, M.E.  1998,
\textit{Mon. Not. R. Astron. Soc.} 297, 1227

 \bibitem[]{}  Hsieh, H.H. \& Jewitt, D. 2006,    
in: D. Lazzaro, S. Ferraz-Mello, \& J.A. Fernandez (eds.),
\textit{Asteroids, Comets, and Meteors}, IAU Symposium 229, 
Cambridge University Press: Cambridge, p. 425 

\bibitem[Ipatov (1987)]{Ipatov1987}
{Ipatov, S.I.} 1987, \textit{Earth, Moon, and Planets} 39, 101

\bibitem[Ipatov (1993)]{Ipatov1993}
{Ipatov, S.I.} 1993, \textit{Solar System Research} 27, 65

\bibitem[Ipatov (1995)]{Ipatov1995}
{Ipatov, S.I.} 1995, \textit{Solar System Research} 29, 261

  \bibitem[]{} Ipatov, S.I. 1999, 
\textit{Celest. Mech. Dyn. Astr.} 73, 107
      
 \bibitem[]{} Ipatov, S.I. 2000, \textit{Migration of 
 celestial bodies in the solar system.}
Editorial URSS Publishing Company: Moscow, 320 p. (in Russian)

\bibitem[Ipatov (2001)]{Ipatov2001}
{Ipatov, S.I.} 2001, \textit{Adv. Space Res.} 28, 1107

  \bibitem[]{} Ipatov, S.I. 2002,  in:  Warmbein, B. (ed.),
\textit{Asteroids, comets, meteors, 2002}, 
SP-500, European Space Agency, 371 

\bibitem[Ipatov (2004)]{Ipatov2004}
{Ipatov, S.I.} 2004, in: S.S. Holt and D. Deming (eds.),
\textit{The Search for Other Worlds}, American Inst. of Physics,
AIP Conference Proceedings, vol. 713, p. 277

 \bibitem[]{}  Ipatov, S.I. \& Hahn, G.J. 1999, 
\textit{Solar System Research} 33,  487

\bibitem[Ipatov \& Mather (2003)]{IpatovM2003} 
{Ipatov S.I. \& Mather J.C.} 2003, \textit{Earth, Moon, \& Planets} 92, 89 

\bibitem[Ipatov \& Mather (2004a)]{IpatovM2004a}
{Ipatov, S.I. \& Mather, J.C.}  2004a, in: E. Belbruno, D. Folta, \& P. Gurfil (eds.),  
\textit{Astrodynamics, Space Missions, and Chaos}, Annals of the New York Acad. of Sci.,
vol. 1017, 46 

\bibitem[Ipatov \& Mather (2004b)]{IpatovM2004b}
{Ipatov, S.I. \& Mather, J.C.} 2004b, \textit{Adv. Space Res.} 33, 1524

 \bibitem[]{}       Ipatov, S. I. \& Mather, J. C. 2006a, 
\textit{Adv. Space Res.}, 37, 126 

 \bibitem[]{}     Ipatov, S. I. \& Mather, J. C. 2006b, in: H. Kruger \& A. Graps (eds.),
\textit{Dust in Planetary System} (ESA),
in press. Available from: http://arXiv.org/format/astro-ph/0606434

\bibitem[Ipatov et al. (2004)]{IpatovMT2004} {Ipatov, S.I., Mather, J.C., \& Taylor, P.A.}   2004, in:
B. Belbruno, D. Folta, \& P. Gurfil  (eds.),
\textit{Astrodynamics, Space Missions, and Chaos}, 
Annals of the New York Acad. of Sci.  (New York: NYAS), vol. 1017,  66

\bibitem[Ipatov et al. (2006)]{Ipatovl2006} {Ipatov, S.I., Kutyrev,  A.S., Madsen, G.J., 
Mather, J.C., Moseley, S.H., Reynolds, R.J.} 2006a, \textit{37th LPSC}, \# 1471 

\bibitem[Ipatov et al. (2006)]{Ipatovo2006} {Ipatov, S.I., Kutyrev,  A.S., Madsen, G.J., 
Mather, J.C., Moseley, S.H., Reynolds, R.J.} 2006b, \textit{AJ}, submitted. 
Available from: http://arXiv.org/format/astro-ph/0608141

  \bibitem[]{} Jewitt, D.  2004, 
in: M.C. Festou, H.U. Keller, \& H.A. Weaver (eds.),
      \textit{Comets II}, The University of Arizona Press,  p. 659
   
  \bibitem[]{} Jewitt, D. \& Fernandez, Y. 2001, 
in: M. Ya. Marov \& H. Rickman (eds.),
      \textit{Collisional Processes in the Solar System}, ASSL, vol. 261, p. 143
   
 \bibitem[]{} Kuchner, M.J., Brown, M.E., \& Holman, M. 2002, 
\textit{Astron. J.}, 124, 1221
   
 \bibitem[]{} Kuchner, M.J., Youdin, A., \& Bate, M. 2004,  
\textit{Proc. of the Second TPF/Darwin International Conference}, 
Available from: http://planetquest1.jpl.nasa.gov/TPFDarwinConf/confProceedings.cfm  

\bibitem[Levison and Duncan (1994)]{Levison1994}  
{Levison, H.F. \& Duncan, M.J.} 1994, \textit{Icarus} 108, 18

\bibitem[Levison and Duncan (1997)]{Levison1997}
{Levison, H.F. \& Duncan, M.J.} 1997, \textit{Icarus} 127, 13

 \bibitem[]{} Levison, H.F., Dones, L., Chapman, C.R., et. al. 2001, 
\textit{Icarus} 151, 286
      
\bibitem[Levison et al. (2006)]{Levison2006}
{Levison, H.F., Terrel, D., Wiegent, P.A., Dones, L., \& Duncan, M.J.} 
2006, \textit{Icarus} 182, 161

 \bibitem[]{} Lunine, J.I. 2004,  
\textit{Proc. of the Second TPF/Darwin International Conference}, 
Available from: http://planetquest1.jpl.nasa.gov/TPFDarwinConf/confProceedings.cfm  

\bibitem []{} Lunine, J.I., Chambers, J., Morbidelli, A., \& Leshin, L.A.
2003, \textit{Icarus} 165, 1

 \bibitem[]{} Lunine, J.I. 2006,   in D.S. Lauretta \& H.Y. McSween Jr. (eds.),
\textit{Meteorites and the Early Solar System II}, 
Univ. of Arizona Press: Tucson, p. 309

 \bibitem[]{} Lupishko, D.F. \& Lupishko, T.A. 2001, 
\textit{Solar System Research} 35, 227
      
\bibitem[]{} {Marov, M.Ya. \& Ipatov, S.I.}  2001, 
in: M. Ya. Marov \& H. Rickman (eds.),
      \textit{Collisional Processes in the Solar System}, ASSL, vol. 261, p. 223

 \bibitem[]{} Marov, M.Ya. \&  Ipatov, S.I. 2005,
 \textit{Solar System Research} 39, 374   

 \bibitem[]{} Merline, W.J., Weidenschilling, S.J., Durda, D.D.,
Margot, J.-L., Pravec, P., \& Storrs, A.D. 2002,
in: W.F. Bottke Jr., A. Cellino, P. Paolicchi, \& R.P. Binzel (eds.),
\textit{Asteroids III}, Univ. of Arizona: Tucson, p. 289 
      
 \bibitem[]{} Morbidelli, A., Chambers, J., Lunine, J.I., Petit, J.M., 
Robert, F., Valsecchi, G.B., \& Cyr, K.E. 2000, 
\textit {Meteoritics \& Planetary Science} 35, 1309 

 \bibitem[]{}      
Noll, K.S. 2006, 
in: D. Lazzaro, S. Ferraz-Mello, \& J.A. Fernandez (eds.),
\textit{Asteroids, Comets, and Meteors}, IAU Symposium 229, 
Cambridge University Press: Cambridge, p. 301.
      
 \bibitem[]{} Pavlov, A.A., Pavlov, A.K., \& Kasting, J.F. 1999, 
\textit{Journal of Geophysical research} 104, No. E12, 30,725
      
 \bibitem[]{} Pravec, P., Scheirich, P., Kusnirak, P. et. al. 2006, 
\textit{Icarus} 181, 63

  \bibitem[]{} Rickman, H., Fernandez, J.A., Tancredi, G., \& Licandro, J. 2001, 
in: M.Ya. Marov \& H. Rickman (eds.),
      \textit{Collisional Processes in the Solar System}, 
(ASSL Dordrecht: Kluwer Academic Publishers), vol. 261, 131
      
  \bibitem[]{} Weissman, P.R., Bottke, W.F. Jr., \& Levison, H.F. 2002, 
in: W.F. Bottke Jr., A. Cellino, P. Paolicchi, \& R.P. Binzel (eds.),
\textit{Asteroids III}, Univ. of Arizona: Tucson, p. 669 

\bibitem[Wetherill (1988)]{Wetherill1988}
{Wetherill, G.W.} 1988, \textit{Icarus} 76, 1

\end{thebibliography}
\end{document}